\documentclass[seceq,preprint]{ptptex}
\usepackage{wrapft}
\usepackage{graphicx}

\newcommand{\wtilde}[1]{\widetilde{#1}} 

\newcommand{\slp}{\raise.1ex\hbox{$/$}\kern-.63em\hbox{$p$}}
\newcommand{\slk}{\raise.15ex\hbox{$/$}\kern-.53em\hbox{$k$}}
\newcommand{\slpartial}{\raise.15ex\hbox{$/$}\kern-.53em\hbox{$\partial$}}
\newcommand{\bp}{\mib{p}}
\newcommand{\bk}{\mib{k}}
\newcommand{\ov}{\overline}
\newcommand{\II}{I\hspace{-.1em}I}

\def\<{\langle}
\def\>{\rangle}

\title{%
Effective Potential Study of the Chiral Phase Transition 
in a QCD-Like Theory
}

\author{%
Yoshinori {\sc Hashimoto}$^{1}$, 
Yasuhiko {\sc Tsue}$^2$ and Hirotsugu {\sc Fujii}$^3$
}

\inst{
$^1$Department of Applied Science, Kochi University, Kochi 780-8520, 
Japan\\
$^2$Physics Division, Faculty of Science, Kochi University, Kochi 780-8520, 
Japan\\
$^3$Institute of Physics, University of Tokyo, Komaba, 
Tokyo 153-8902, Japan
}

\recdate{
\today
}

\abst{%
We construct the effective potential for a QCD-like theory using
the auxiliary field method. 
The chiral phase transition exhibited by the model at finite temperature 
and the quark chemical potential
is studied from the viewpoint of the shape change of the potential
near the critical point.
We further generalize the effective potential so as to have
quark number and scalar quark densities as independent variables
near the tri-critical point.
}
\begin{document}
\maketitle



\section{Introduction}

The phase structure of quantum chromodynamics (QCD) 
at finite temperature $T$ 
and quark chemical potential $\mu$ has 
has been investigated since the invent of QCD. 
\cite{CP75}
Today collider experiments using ultra-relativistic heavy ion
beams are creating highly-excited QCD matter in the 
laboratory.\cite{QGP} 
Extremely dense, cold hadronic matter is relevant to the
physics of the inner structure of neutron stars.\cite{KMTT93}

It is currently an accepted concept that the 
chiral symmetry of QCD, which is spontaneously broken in the vacuum, 
will be restored at sufficiently high temperature 
and/or quark chemical potential.
Although steady progress has been made in the study of lattice QCD at finite 
$T$ and $\mu$, 
numerical simulations are still hard to be carried out with physically 
realistic conditions, 
especially for finite $\mu$. \cite{LQCDatNara}
Many model studies have been undertaken 
and have provided information concerning the in-medium properties of 
QCD\cite{HK94}. 
In Ref.~\citen{AY89}, the $T$--$\mu$ phase diagram
was investigated with the Nambu-Jona-Lasinio model,
and evidence was found suggesting 
the existence of the tri-critical point.
The phase diagram was also studied using the Schwinger-Dyson
(SD) equation for the fermion propagator with a variational ansatz in 
a QCD-like theory\cite{BCCGP89,BCP94}. 
With the possibility of color superconductivity being newly
suggested in Refs.~\citen{BL,II,RW,RSSV},
the phase structure at high density but low temperature 
has been re-investigated by using the diquark condensate
as a new order parameter, in addition to the usual quark--anti-quark 
$q \bar q$ condensate.

The effective potential is a useful tool, 
and it provides a clear picture regarding the nature of the phase transitions
in these studies.
In this paper 
we re-investigate the chiral phase transition of
a QCD-like theory, 
focusing on the shape change of the 
effective potential near the critical point.

The QCD-like theory\cite{ABKMN90,K92,H91} is the 
renormalization-group (RG) improved ladder approximation for the 
SD equation of QCD.
This model describes
the dynamical chiral symmetry breaking in the vacuum
while retaining the correct high energy behavior of the quark mass
function.
Using this model and the low-density expansion,
the pion-nucleon sigma term and the quark condensate at 
finite density have been calculated.\cite{FT95}. 
The results are consistent with those of other models.
The chiral phase transition at finite $T$ and/or $\mu$
in this model was studied in Refs.~\citen{TY97,HS98,KMT00,I02}. 
The color superconducting phase 
has also been investigated.\cite{T03,A03}

The Cornwall-Jackiw-Tomboulis (CJT) potential functional \cite{CJT74} 
has been used in several works\cite{BCP94,KMT00} attempting to determine 
the non-perturbative extremum solution for the model.
For instance, in Ref.~\citen{KMT00}, 
the mass function was assumed to have a certain trial form
characterized by the value of the scalar $q \bar q$ condensate, 
which was
determined by extremizing the CJT potential variationally.
However, the interpretation of the CJT potential away from the extremum
is not obvious\cite{RWH91}. The solution is known to be
a saddle point of the CJT action with respect to general variations.

Instead of following the approach outlined above, 
we construct an effective potential using the auxiliary
field method in the QCD-like theory.
Within the stationary phase approximation, 
the solution of this potential is known to be a local
minimum,\cite{RWH91} and 
it coincides with the CJT potential when
the external field is turned off.
This is a suitable property for a variational method, and it 
could be applied to other order parameters, such as the diquark
condensate, provided that the gauge invariance is treated properly.

This paper is organized as follows. 
In the next section, we review the QCD-like theory 
in the vacuum and at finite $T$ and $\mu$.
In \S 3, we define the effective potential for the $q\bar q$ condensate
in the QCD-like theory.
Furthermore, we generalize the potential such that it possesses 
the quark number density as the second order parameter,
in addition to the $q \bar q$ condensate.
We investigate 
the transition points through consideration of the global 
properties of these potentials.
The last section is devoted to a summary. 
In Appendix A, we demonstrate the validity of our definition of the 
effective potential in the Nambu-Jona-Lasinio model.

\section{The QCD-like theory}

In this section, 
we briefly review the QCD-like theory \cite{ABKMN90,K92,H91},
introducing the notation and approximations used in this paper.

\subsection{In the vacuum}

Let us start by deriving the SD equation in the QCD-like theory 
as an extremum condition for the effective potential 
obtained using the auxiliary field method at zero
temperature and zero chemical potential.
\cite{ABKMN90,K92,H91}

First, we integrate out the gluon field in the expression 
for the partition function $Z$,
omitting the gluon self-interaction term in the QCD Lagrangian 
with massless quarks:
\begin{eqnarray}\label{1}
Z&\propto&
\int\mathcal{D}\psi\mathcal{D}\overline{\psi}\mathcal{D}A 
\exp i\int d^4x \mathcal{L} 
\nonumber\\
&\rightarrow& 
\int\mathcal{D}\psi\mathcal{D}\overline{\psi} 
\exp \biggl[ i\int_p\overline{\psi}(p)\ \slp\ \psi(p) \nonumber\\ 
& &\ 
-\frac{i}{2}\int_{pqk}  
\psi_\alpha\left(p-\frac{q}{2}\right)
\overline{\psi}_\beta\left(p+\frac{q}{2}\right)
K^{\alpha\beta,\gamma\delta}(p,k) 
\psi_\gamma\left(k+\frac{q}{2}\right)
\ov{\psi}_\delta\left(k-\frac{q}{2}\right)
\biggr] ,\ \ \ 
\end{eqnarray}
where $\psi(p)$ is the quark field in the momentum space 
and $\int_p\equiv\int d^4p/(2\pi)^4$.
The indices, $\alpha, \beta, \cdots$ correspond to the Dirac structure, and 
$T^a$ is the color $su(N_c)$ generator. 
Here, we defined the kernel $K$ as
\begin{equation}\label{2}
K^{\alpha\beta,\gamma\delta}(p,k)=g^2(\gamma_\mu T^a)^{\delta\alpha}
(\gamma_\nu T^a)^{\beta\gamma}iD^{\mu\nu}(p-k) \ ,
\end{equation}
with the gluon propagator $iD^{\mu\nu}(p)$ given by 
\begin{equation}\label{3}
iD^{\mu\nu}(p)=\frac{g^{\mu\nu}-(1-\alpha)p^\mu p^\nu / p^2}{p^2}.
\end{equation}
The kernel $K$ represents the gluon exchange between the quarks.
Hereafter, we employ the Landau gauge ($i.e.$, $\alpha=0$).
The non-Abelian nature of the gluon interaction is treated in this model
as the one-loop running of $g$ in (\ref{2}):
$\ov{g}(\textrm{max}(p_E^2,k_E^2))$, \cite{HM84}
with $p_E$ the momentum in Euclidian space.
The divergence of $\ov{g}(p_E^2)$ appearing 
at $p_E=\Lambda_{\rm QCD}$, 
is removed by introducing an infrared 
cutoff parameter $p_{IF}$\cite{HM84,sasaki} as 
\begin{equation}\label{4}
\overline{g}^2(p_E^2)
=\frac{2}{a}\frac{1}{\textrm{ln}((p_E^2+p^2_{IF})/\Lambda_{\rm QCD}^2)} \ ,
\end{equation}
with
\begin{equation}
a=\frac{1}{8\pi^2}\frac{11N_c-2N_f}{3} .
\end{equation}
Here, 
$N_c$ and $N_f$ are the numbers of colors and flavors, respectively.

We next introduce the following bilocal auxiliary field 
for non-perturbative analysis:
\begin{equation}\label{5}
\chi_{\alpha\beta}(p,q)=\int_k K^{\alpha\beta,\gamma\delta}(p,k)
\psi_\gamma\left(k+\frac{q}{2}\right)
\ov{\psi}_\delta\left(k-\frac{q}{2}\right).
\end{equation}
This makes the action bilinear in the quark fields $\psi$
and $\ov{\psi}$.
Then, integrating out $\psi$ and
$\ov{\psi}$, we obtain the classical action 
$Z=\int {\cal D}\chi \exp i\,\Gamma[\chi]$ as
\begin{eqnarray}\label{6}
\Gamma[\chi]&=&
\frac{1}{2}\int_{pkq}\textrm{tr}[\chi(p,-q)K^{-1}(p,k)\chi(k,p)] 
\nonumber\\ 
& &
\qquad
-i\textrm{Tr Ln}(\ \slp\delta^4(q)(2\pi)^4-\chi(p,-q))  .
\end{eqnarray}
The SD equation is obtained
as the extremum condition for this classical action
within the stationary-phase approximation for $\chi$.\cite{ABKMN90}
The existence of a non-trivial solution for $\chi$ indicates 
the dynamical breaking of the
chiral symmetry.

Due to the translational invariance of the vacuum,
this solution is expressed as
$\langle \chi_{\alpha\beta} \rangle 
=\Sigma_{\alpha\beta}(p)\delta^4(q)(2\pi)^4$, 
with the mass function $\Sigma$, and 
the effective potential $V[\Sigma]=-\Gamma[\chi]/\int d^4x$ 
becomes
\begin{eqnarray}\label{7}
V[\Sigma]&=&
-\frac{1}{2}\int_{pk}\textrm{tr}[\Sigma(p)K^{-1}(p,k)\Sigma(k)] 
+i\int_p\textrm{tr ln}(\ \slp-\Sigma(p)) .
\end{eqnarray}
Then the SD equation, $\delta V/\delta \Sigma=0$, 
in the improved ladder approximation is written 
\begin{equation}\label{8}
\Sigma_{\alpha\beta}(p)
=\frac{1}{i} \int_k K^{\alpha\beta,\gamma\delta}(p,k) 
\biggl(\frac{1}{\slk -\Sigma(k)} \biggr)_{\gamma\delta} .
\end{equation}
Whereas $\Sigma_{\alpha\beta}(p)$ has the general form
 $\Sigma(p^2)\delta_{\alpha\beta}+\Sigma_v(p^2) \slp_{\alpha\beta}$
in the vacuum,
the $\Sigma_v(p^2)$ part is known to vanish 
in the Landau gauge ($\alpha=0$) \cite{MN74}. 
After carrying out the Wick rotation and the angle integration, 
the SD equation becomes
\begin{eqnarray}\label{9}
\Sigma(p^2_E)&=&
\frac{3C_2(N_c)}{16\pi^2}\int_0^{\infty} k_E^2 
dk_E^2 \overline{g}^2(\textrm{max}(p_E^2,k_E^2)) 
\frac{1}{\textrm{max}(p_E^2,k_E^2)}
\frac{\Sigma(k^2_E)}{k_E^2+\Sigma(k^2_E)^2} , \qquad
\end{eqnarray}
where $C_2(N_c)=T^aT^a=(N_c^2-1)/2N_c$.

\subsection{Order parameters of the chiral symmetry}

The quark condensate with the four momentum cutoff $\Lambda$ is defined as  
\begin{eqnarray}\label{10}
\langle \overline{\psi}\psi \rangle_\Lambda&=&
-\frac{1}{i}\int_p\textrm{tr}\left(\frac{1}{\slp-\Sigma(k^2)}\right) 
\nonumber\\
&=&-\frac{N_c}{4\pi^2}\int^{\Lambda^2}_0 \!\! dp_E^2 \ 
\frac{p_E^2\Sigma(p_E^2)}{p_E^2+\Sigma(p_E^2)^2} . 
\end{eqnarray}
This bare value at the scale $\Lambda$  is converted into 
the value at the lower energy scale $\mu$ (e.g., 1 GeV) 
via the renormalization group equation 
\begin{equation}\label{11}
\langle \overline{\psi}\psi \rangle_{\mu}
= \langle \overline{\psi}\psi \rangle_\Lambda
\left(\frac{\ov{g}^2(\Lambda)}{\ov{g}^2(\mu)}\right)^\frac{1}{4B} , 
\end{equation}
where $B=(12C_2(N_c))^{-1}(11N_c-2N_f)/3.$
We will present the value of $\langle \overline{\psi}\psi \rangle_{\mu}$ at
$\mu=1$ GeV and omit the subscript $\mu$ hereafter.

Next, the pion decay constant $f_\pi$
is estimated in terms of the mass function 
$\Sigma(p^2)$ by utilizing the Pagels-Stokar formula: \cite{pagels}
\begin{equation}\label{12}
f^2_\pi=\frac{N_c}{4\pi^2}\!
\int_0^{\infty}\!\!\!
 dp_E^2 \frac{p_E^2\Sigma(p_E^2)}{(p_E^2+\Sigma^2(p_E^2))^2}\!
\left(\!\!\Sigma(p_E^2)\!-\!
\frac{p_E^2}{2}\frac{d\Sigma(p_E^2)}{dp_E^2}\!\!\right) .
\end{equation}
We fix the value of $\Lambda_{\rm QCD}$ 
here 
so as to reproduce the empirical value of $f_\pi$ 
with this formula.

\subsection{At finite temperature and chemical potential}

We use the imaginary time formalism to extend the QCD-like
theory to the case with finite temperature and chemical potential, 
making the following replacement:~\footnote{We ignore 
the possible dependence of 
the running coupling constant $\ov{g}^2$
on the scales, $T$ and/or $\mu$.
\cite{KKZP04}.}
\begin{equation}\label{13}
\int_p f(p_0,\bp) \longrightarrow 
T\sum_{n=-\infty}^\infty \int\frac{d^3\bp}{(2\pi)^3}f(i\omega_n+\mu,\bp) , 
\end{equation}
where $\omega_n=(2n+1)\pi T \; (n \in \mathbb{Z} )$ is the Matsubara 
frequency for the fermion.

The mass function $\Sigma_{\alpha\beta}(\omega_n,\mib{p})$ 
at finite $T$ and $\mu$,
is invariant under spatial $O(3)$ rotations and 
decomposes into 
$\Sigma(\omega_n,|\mib{p}|)\delta_{\alpha\beta}
+\Sigma_s(\omega_n,|\mib{p}|)\omega_n(\gamma_0)_{\alpha\beta}
+\Sigma_v(\omega_n,|\mib{p}|)p^i(\gamma_i)_{\alpha\beta}$. 
It is in fact possible\cite{I02}, although still cumbersome, 
to solve the SD equation numerically in this general form.
For our purpose of demonstrating the usefulness of the effective
potential in the QCD-like theory,
we assume here $\Sigma_s=\Sigma_v=0$ for simplicity. 
We, further, use a covariant-like ansatz\cite{sasaki} for the mass function,
taking 
\begin{eqnarray}\label{14}
& &\Sigma(\omega_n,\bp) \longrightarrow \Sigma(\hat{p}^2) , 
\end{eqnarray}
where the frequency and the momentum 
appear in the combination $\hat{p}^2=\omega_n^2+|\bp|^2 $
in $\Sigma$.
Thus, the SD equation simplifies to
\begin{eqnarray}\label{15}
\Sigma(\hat{p}^2)&=&
\frac{3C_2(N_c)}{8\pi^2} T\!\! 
\sum_{m=-\infty}^\infty \int_{\omega_m^2}^\infty 
d\hat{k}^2\frac{\ov{g}^2(\hat{p}^2,\hat{k}^2)}{\sqrt{\hat{p}^2-w_n^2}} 
\nonumber\\ 
& &\times \textrm{ln}\left[\frac{\hat{p}^2+\hat{k}^2+2\sqrt{(\hat{p}^2-w_n^2)
(\hat{k}^2-w_m^2)}-2w_nw_m}
{\hat{p}^2+\hat{k}^2-2\sqrt{(\hat{p}^2-w_n^2)(\hat{k}^2-w_m^2)}-2w_nw_m}\right] \nonumber\\ 
& &\qquad\times 
\frac{\Sigma(\hat{k}^2)}{\hat{k}^2+2i\mu\omega_m-\mu^2
+\Sigma(\hat{k}^2)^2} .
\end{eqnarray}
At finite chemical potential, solutions of this equation are
generally complex-valued.
We set $\omega_n=0$ on the right-hand side of Eq.~(\ref{15}),
which makes the SD equation real.
At high temperature ($T \gg |\bp|$ and $T \gg |\bk|$), 
the Matsubara frequencies with large $n$ 
give only small corrections to physical quantities.\cite{sasaki}

Substituting the solution $\Sigma$ into the 
effective potential (\ref{7}) with the replacement (\ref{13}), 
the extremum value $V_{\rm{ex}}$ reads
\begin{eqnarray}\label{17}
V_{\rm{ex}}[\Sigma]&=&
\frac{N_cN_f}{2\pi^2}T\sum_{n=0}^\infty \int_{\omega_n^2}^\infty d\hat{p}^2 
\, \sqrt{\hat{p}^2-\omega_n^2} \nonumber\\
& &\times 
\left[2 \frac{(\hat{p}^2-\mu^2+\Sigma^2(\hat{p}^2))\Sigma^2(\hat{p}^2)}
{(\hat{p}^2-\mu^2+\Sigma^2(\hat{p}))^2+4\mu^2\omega_n^2} \right. 
- \ln[(\hat{p}^2-\mu^2+\Sigma^2(\hat{p}^2))^2+4\mu^2\omega_n^2] \biggr] .
\nonumber\\
& &
\end{eqnarray}
The stability of the symmetry-broken phase is usually examined by 
comparing this extremum value with that of the trivial solution.
It is preferable to have a functional form of the
effective potential in order to study the features of the phase transition. 
In the next section, we present a method for determining the shape of
the effective potential for the composite field.

\section{Shape of the effective potential}

\subsection{The effective potential away from the extremum}

Here we explain a method that we use
to construct the functional form of the effective potential numerically.
A standard way to assess the potential form is to apply an external
source which is coupled to the field linearly.
It is recognized, however, that using this approach 
we cannot study non-convex potentials,
which is an important feature near the phase transition point.
 
We apply an external source field $J(p,k)$ which is
coupled to the square of the self-energy as 
\begin{equation}\label{18}
\widetilde{V}[\Sigma,J]
=V[\Sigma]+\frac{1}{2}\int_{pk}\textrm{tr}[\Sigma(p)J(p,k)\Sigma(k)] ,
\end{equation}
where $V[\Sigma]$ has been defined in (\ref{7}). 
By imposing the extremum condition for ${\wtilde V}[\Sigma, J]$ 
with respect to 
$\Sigma(p)$ (that is, $\delta\wtilde{V}[\Sigma,J]/\delta \Sigma(p)=0$), 
we derive the SD equation with the source field $J$ as 
\begin{equation}\label{19}
\Sigma(p)=\frac{1}{i}\int_k (K^{-1}-J)^{-1}(p,k)\frac{1}{\slk-\Sigma(k)} \ .
\end{equation}
We denote the solution of (\ref{19}) by $\Sigma_J(p)$ to indicate the
implicit dependence on the source $J$.
The effective potential $V$ for the configuration
$\Sigma_J(p)$ is written 
\begin{eqnarray}\label{20}
{V}[\Sigma_J]
&=&{\widetilde V}[\Sigma_J,J]
-\frac{1}{2}\int_{pk}\textrm{tr}[\Sigma_J(p)J(p,k)\Sigma_J(k)]
\nonumber\\
&=&
-\frac{1}{2} \int_p \textrm{tr}
\left [ \Sigma_J(p)\frac{-i}{\slp-\Sigma_J(p)} \right ]
+i\int_p {\rm tr} \ln(\slp-\Sigma_J(p))
\nonumber\\
& &-\frac{1}{2}\int_{pk}\textrm{tr}[\Sigma_J(p)J(p,k)\Sigma_J(k)] .
\end{eqnarray}
We study the global behavior of $V[\Sigma_J]$ near the critical points
by varying $\Sigma_J$ through a particular type of variation of a 
function $J(p,k)$.

Here, among infinite possibilities, we consider one natural choice, 
$J(p,k)= -c K^{-1}(p,k)$, with a parameter $c$,
and study the shape of the effective potential along this particular variation.
Then, the SD equation becomes
\begin{equation}\label{21}
\Sigma_c(p)=\frac{-i}{1+c} \int_k K(p,k)\frac{1}{\slk-\Sigma_c(k)} .
\end{equation}
We note that the variation of $c$ is, in effect, equivalent
to varying the strength of the strong coupling constant, 
since
$(K^{-1}-J)^{-1} = K/(1+c) \propto {\bar g}^{2}/(1+c)$. 
In the limit $c\rightarrow \infty$ we should have a trivial solution, 
while a symmetry-breaking solution, $\Sigma_c \ne 0$, exists
in the case $c\rightarrow -1$, which
corresponds to the case of infinite effective coupling.\footnote{
This situation is similar to that in the strong coupling QED investigated 
previously\cite{morozumi}.}\footnote{For this particular choice for $J$, 
one could also utilize
the Hellmann-Feynman theorem.
We acknowledge referee's comment on this point.}
In this way, we should be able to scan the potential shape 
along a family of 
configurations, from a trivial one to a symmetry-breaking one.

By substituting the solution $\Sigma_c(p)$  of (\ref{21}) 
into $V[\Sigma_c]$ in (\ref{20}), 
we obtain the effective potential as a function of $c$:
\begin{eqnarray}\label{22}
V(c)&=&
\frac{1}{2}\frac{i}{1+c}\int_p
\textrm{tr}\left [\Sigma_c(p)\frac{1}{\slp-\Sigma_c(p)}\right ]
+i\int_p\textrm{tr ln}( \slp-\Sigma_c(p)).
\end{eqnarray}
As the scalar condensate $\langle \bar \psi \psi \rangle$ is
computed with $\Sigma_c$, the effective potential can be
expressed numerically as a function of $\langle \bar \psi \psi \rangle$.
Near a first-order transition point, the SD equation (\ref{21})
has two non-trivial solutions for a certain range of $c$, each of which 
corresponds to a different value of the scalar condensate.
Using both the solutions, we obtain the effective potential
for the full range of the scalar condensate.
It is straightforward to extend this potential to a system at 
finite temperature and chemical potential
using the replacement expressed in (\ref{13}).

In the numerical evaluation of the
effective potential,
we split $V[\Sigma]$ as  
\begin{equation}\label{25}
V[\Sigma]= (V[\Sigma]-V[0]) + V[0] .
\end{equation}
Here, 
the difference between the free energies $V[\Sigma_J]-V[0]$ is
finite and can be evaluated numerically. 
$V[0]=i\int_p\textrm{tr ln}\ \slp $
 is 
the potential of a free massless quark gas 
and is 
divergent due to the vacuum fluctuations.
It is thus necessary to adopt an appropriate regularization to 
remove this divergence.
In our approximation, we replace 
the last term $V[0]$ in (\ref{25})
with the pressure  $P_{\rm{free}}$ of a free massless
quark gas:
\begin{equation}\label{26}
-V[0]\to  P_{\rm{free}} = 
N_cN_fT^4\left[\frac{7\pi^2}{180}+\frac{1}{6}\left(\frac{\mu}{T}\right)^2
+\frac{1}{12\pi^2}\left(\frac{\mu}{T}\right)^4 \right]  .
\end{equation}

In studying the nature of the phase transition, 
especially near a tri-critical point,
it would be instructive to 
extend the effective potential to a form with 
two independent order parameters, 
the quark number density and the scalar density.\cite{FO04}\footnote{In the 
case of the QCD critical point with 
a finite current mass, this extension of the 
potential becomes more useful.}
To this end, we first replace the variable $\mu$ with 
the quark number density $\rho$
through the usual Legendre transformation, 
\begin{equation}\label{31}
F(\langle \bar \psi \psi \rangle,\rho, T)
=V(\langle \bar \psi \psi\rangle,\mu,T)+\rho\mu \ ,
\end{equation}
with
\begin{eqnarray}\label{29}
\rho&\equiv&
-\frac{\partial \, V}{\partial \, \mu} (\langle \bar \psi \psi \rangle,\mu,T) 
\ . 
\end{eqnarray}
Then, by adding a coupling energy with an external potential $\nu$ 
to Eq.~(\ref{31}), 
we define a new effective potential as
\begin{equation}\label{32}
\ov{V}(\langle \bar \psi \psi\rangle, \rho ;\nu,T)
=F(\langle \bar \psi \psi \rangle,\rho, T)-\rho \nu ,
\end{equation}
which may be interpreted as a Landau potential.
\cite{NG92}
In order to construct a non-convex function of $\rho$ we use the unstable
solution of the SD equation.
Note that $\rho$ and $\nu$ are independent variables here.
When
the condition
$\partial \ov{V}/\partial \rho = \mu-\nu = 0$ is imposed,
this new potential  $\ov V$
coincides with the original potential:
$\ov{V}(\langle \bar \psi \psi\rangle, \rho(\mu) ;\mu,T)
=V(\langle \bar \psi \psi\rangle ;\mu,T)$.

\begin{figure}[t]
\begin{center}
\includegraphics[height=4.8cm]{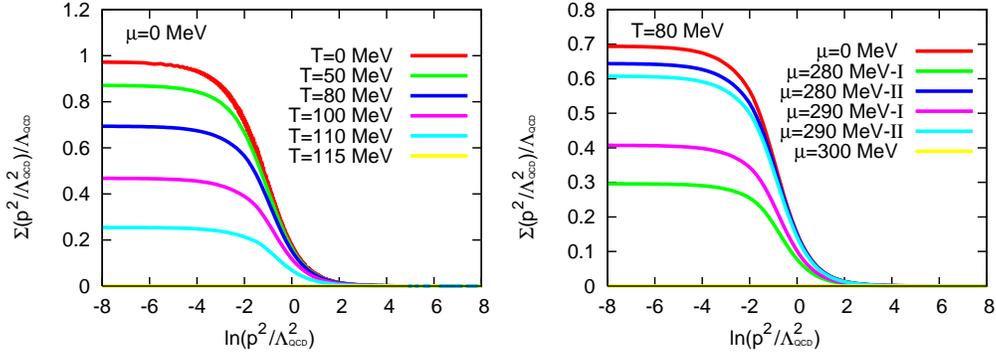}
\caption{The mass functions $\Sigma(p)$ as functions of $p$. 
The left panel displays $\Sigma(p)$ with $\mu=0$ MeV at 
$T=0$, $50$, $80$, $100$, $110$ and $115$ MeV, respectively. 
The right panel displays $\Sigma(p)$ with $T=80$ MeV 
at $\mu=0$, $280$, $290$ and $300$ MeV, respectively. } 
\label{fig:1}
\end{center}
\end{figure}
\begin{figure}[t]
\begin{center}
\includegraphics[height=7.9cm]{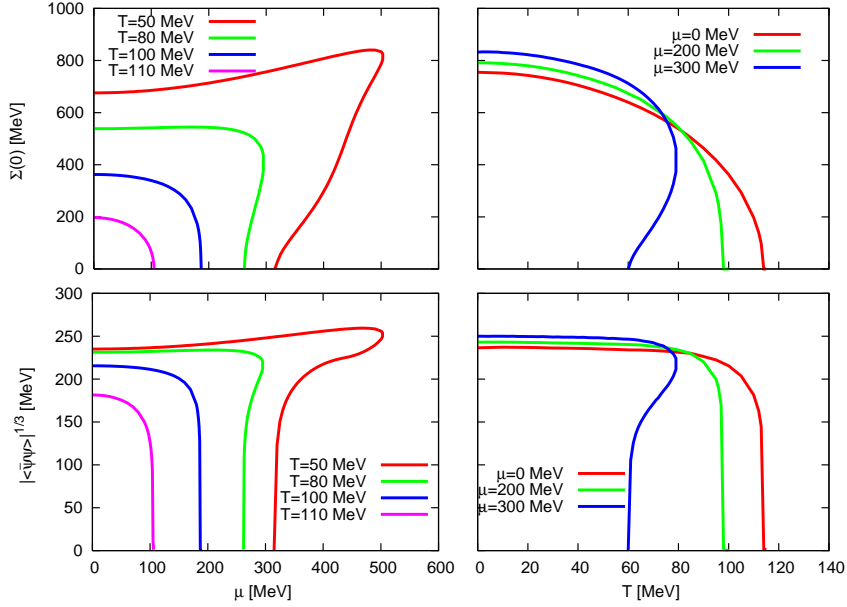}
\caption{The chemical potential (left) and temperature (right) dependences 
of 
$\Sigma(0)$ and $|\langle \overline{\psi}\psi \rangle|^{1/3}$, 
respectively. 
The left figures plot these quantities at 
$T = 50, 80, 100$ and $110$ MeV as functions of $\mu$. 
The right figures plot these two quantities at 
$\mu = 0, 200$ and $300$ MeV as functions of $T$. 
}
\label{fig:2}
\end{center}
\end{figure}

\subsection{Numerical results}

We studied the case with
two massless quark flavors ($i.e.$, $N_f=2$) 
and $N_c=3$, where $N_c$ is the number of colors. 
The model parameter $\Lambda_{\rm QCD}$ 
is fixed 
so as to reproduce the pion decay constant $f_\pi=93$ MeV in (\ref{12}),
and we obtain $\Lambda_{\rm QCD} \simeq 776 $ MeV using
the IR cutoff at $\ln(p_{IF}^2/\Lambda_{\rm QCD}^2)=0.1$.
Then, the quark condensate in the vacuum is found to be
$|\langle \overline{\psi}\psi \rangle|^{1/3} \simeq 236$ MeV 
in this model, and it is insensitive to the value of the UV cutoff. 
(We use $\Lambda^2/\Lambda_{\rm QCD}^2=2.0\times 10^4$.)

At finite temperature and density, we numerically calculated 
the quark mass function $\Sigma(p)$ and the quark condensate 
$\langle \bar \psi \psi \rangle$. 
In Fig.~\ref{fig:1} we plot the mass function $\Sigma(p)$.
The left panel displays 
$\Sigma(p)$ with $\mu=0$ MeV at $T=0$, $50$, 80, $100$, $110$ and $115$ MeV, 
respectively,
while the right panel displays $\Sigma(p)$ with fixed $T=80$ MeV at 
$\mu=0$, $280$, $290$ and $300$ MeV, respectively. 
At the finite chemical potentials
$\mu=280$ and $290$ MeV with $T=80$ MeV, 
there exist two non-trivial solutions for $\Sigma(p)$, which 
we denote I and \II.
The free energy of the solution I is lower than that of
\II.

In the left panels of Fig.~\ref{fig:2}, we plot the $\mu$ dependences
of $\Sigma(0)$ and $\langle \overline{\psi}\psi \rangle$, 
respectively,
at $T = 50, 80, 100$ and $110$ MeV.
The right panels present the $T$ dependences of the same quantities 
at $\mu = 0, 200$ and $300$ MeV.
In the region of low temperature and large chemical potential,
we find two non-trivial extremum solutions, $\Sigma\ne 0$, 
in addition to 
the trivial one, which indicates a first-order transition.
It should be noted here that the chiral condensate $\langle {\bar \psi}
\psi\rangle$ and the value of $\Sigma(0)$ increase as $\mu$ increases 
in the case of fixed $T=50$ MeV, as is seen in Fig.~\ref{fig:2}. 
This behavior is known to be an artifact of 
the approximation made by omitting $\Sigma_s$ and $\Sigma_v$. 
If $\Sigma_s$ and $\Sigma_v$ are properly taken into account, 
$\langle {\bar \psi}\psi\rangle$ and 
$\Sigma(0)$ are found to decrease monotonically, 
and the chiral symmetry is restored,
as $\mu$ increases.\cite{I02} 
%

\begin{figure}[t]
\begin{center}
\includegraphics[height=4.3cm]{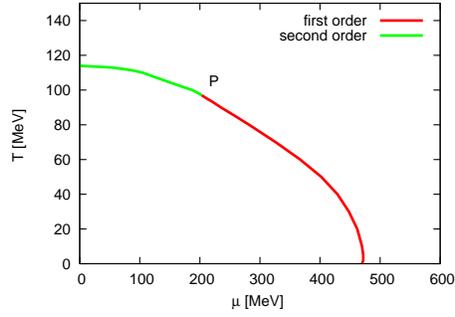} 
\caption{The phase diagram for the chiral symmetry. 
The transition point is located at $T_c \simeq 114$ MeV for $\mu=0$ MeV, 
$\mu_c \simeq 472$ MeV at $T=0$ MeV, and so on. 
The tri-critical point $P$ is located at 
$(T_t, \mu_t)\simeq (97,203)$ MeV. }
\label{fig:7}
\end{center}
\end{figure}
\begin{figure}[t]
\begin{center}
\includegraphics[height=11cm]{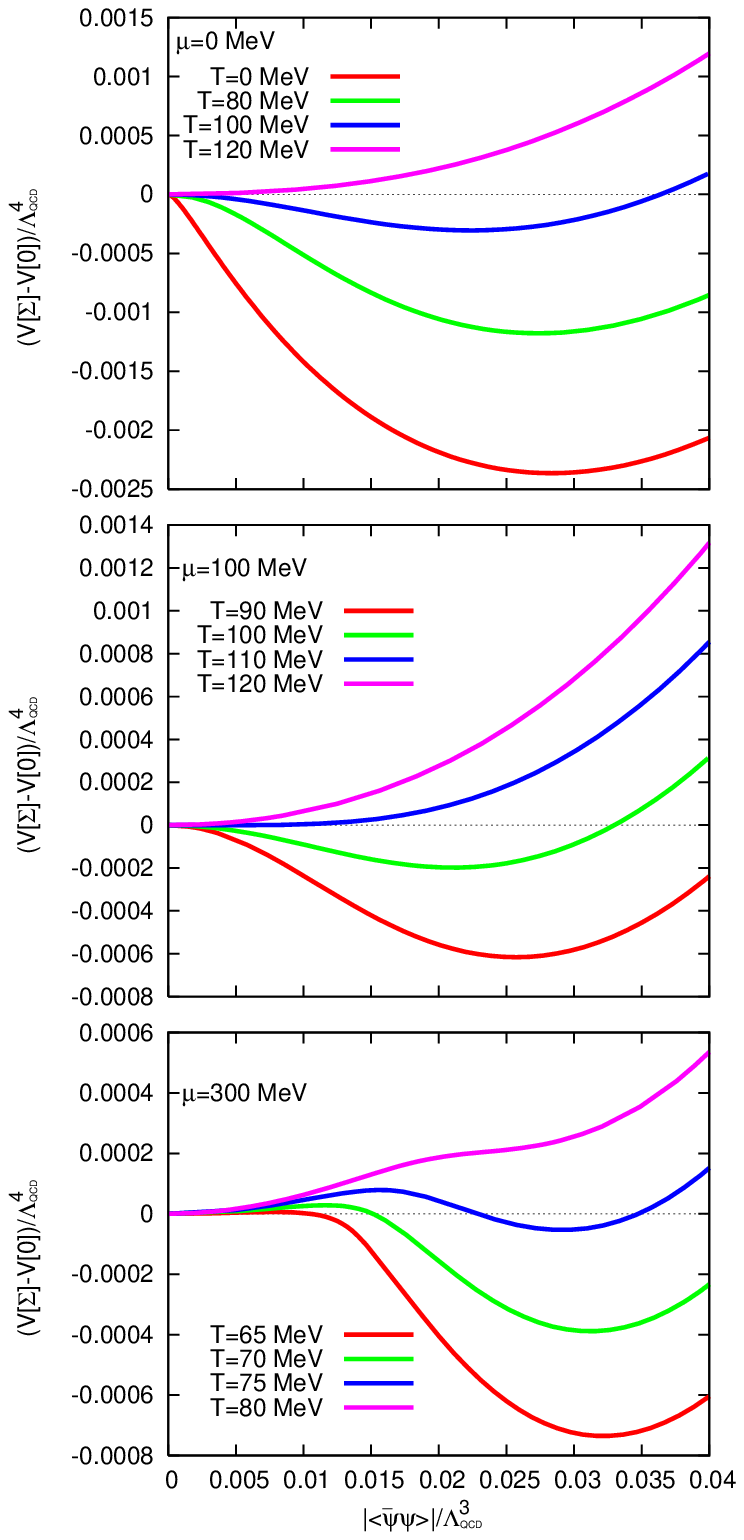}
\caption{The temperature and chemical potential dependence of 
the effective potential, 
$V[\Sigma_c,J]-V[0]$, with fixed $\mu$ as a function of 
$|\langle{\bar \psi}\psi\rangle|$.}
\label{fig:3}
\end{center}
\end{figure}

\begin{figure}[t]
\begin{center}
\includegraphics[height=4.2cm]{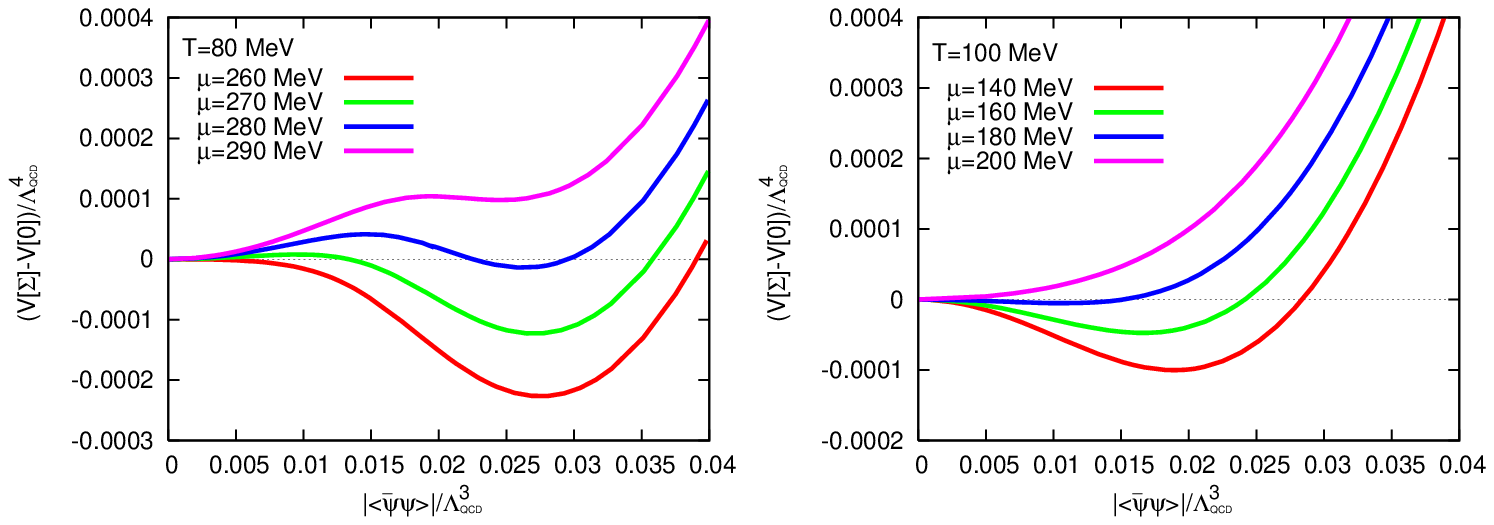}
\caption{The temperature and chemical potential dependence of 
the effective potential, 
$V[\Sigma_c,J]-V[0]$, with fixed $T$ as a function of 
$|\langle{\bar \psi}\psi\rangle|$.}
\label{fig:4}
\end{center}
\end{figure}

\begin{figure}[t]
\begin{center}
\includegraphics[height=4cm]{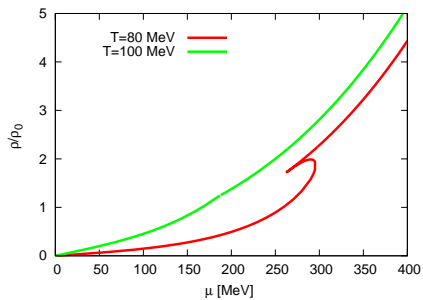}
\caption{The chemical potential dependence of the quark number 
density at $T= 80$ and $100$ MeV. Here, 
$\rho_0=3\times 0.17\ {\textrm{fm}^{-3}}$.
}
\label{fig:5}
\end{center}
\end{figure}
%
\begin{figure}[t]
\begin{center}
\begin{tabular}{cc}
\resizebox{53mm}{!}
{\includegraphics[height=10.4cm]{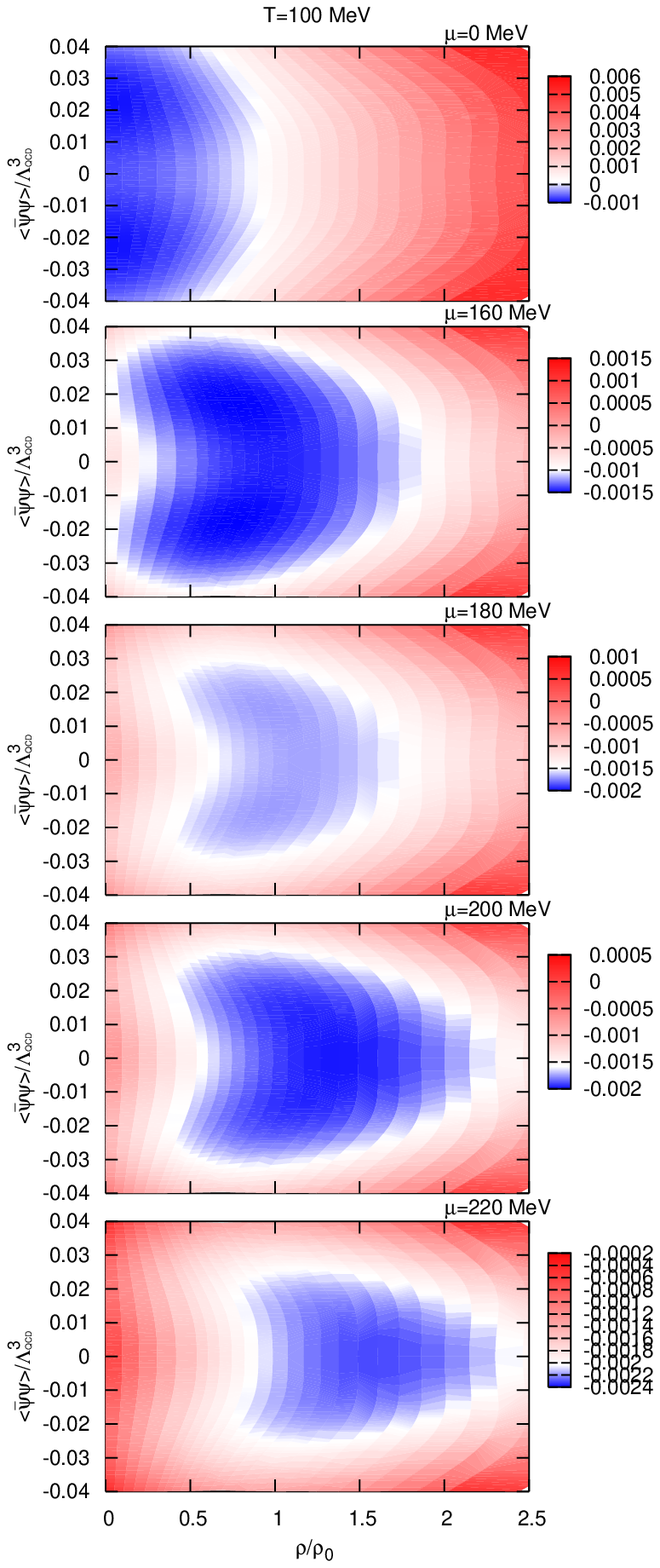}} &
\resizebox{53mm}{!}
{\includegraphics[height=10.4cm]{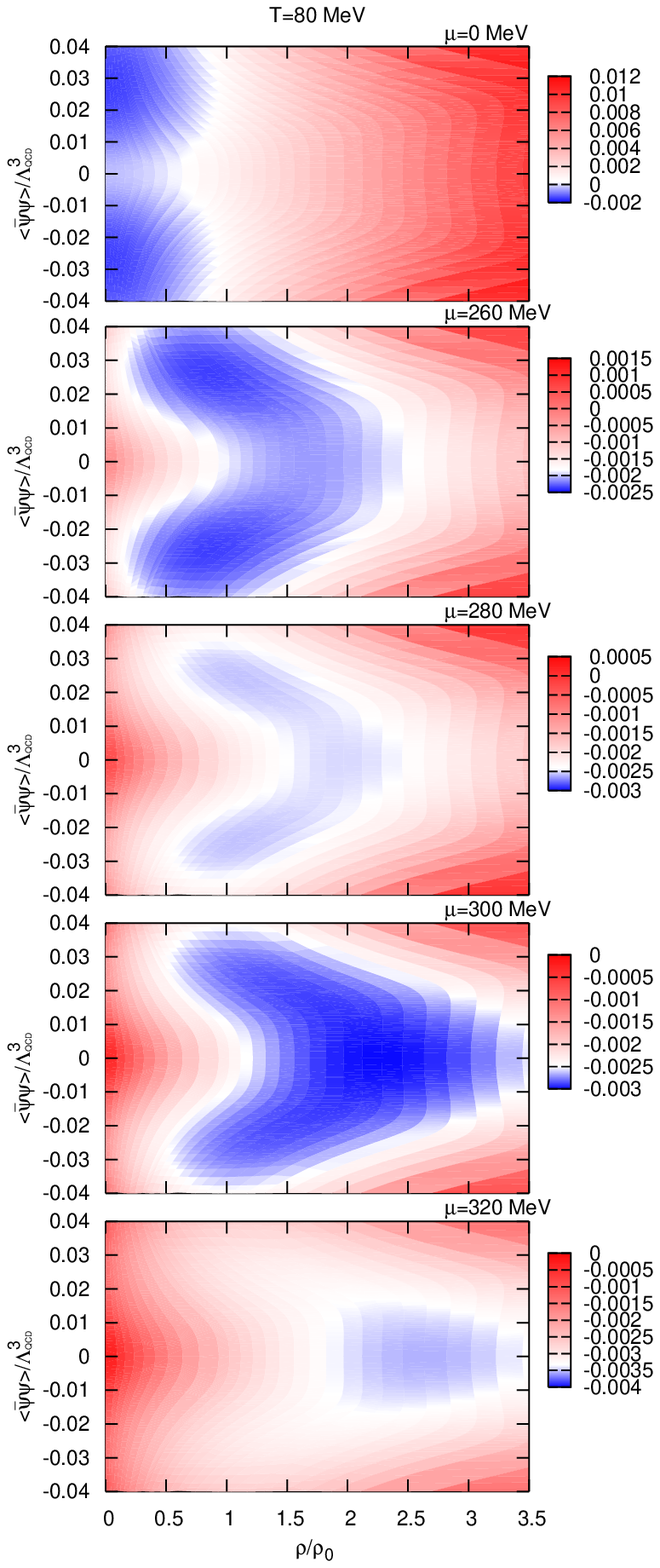}} \\
\end{tabular}
\caption{The contour map for $\ov{V}(\langle{\bar \psi}\psi\rangle,\rho\ ;
\mu, T)$ in the 
$\rho$-$\langle{\bar \psi}\psi\rangle$ plane with various chemical potentials 
for $T=100$ MeV (left) and $T=80$ MeV (right). 
}
\label{fig:6}
\end{center}
\end{figure}

We present the phase diagram of the model in Fig.\ref{fig:7}, where
the tri-critical point appears at $(T_t, \mu_t)\simeq (97 , 203)$ MeV.

The situation may be more clearly understood if we can obtain 
the explicit functional form of the effective potential.
We plot $V[\Sigma_c]-V[0]$ as a function of 
$|\langle {\ov \psi}\psi\rangle|/\Lambda_{\rm QCD}^3$ 
in Fig.~\ref{fig:3} at $\mu=0$, 100 and 300 MeV for several values of $T$.
In Fig.~\ref{fig:4} we plot the results at $T=80$ and 100 MeV with several
values of $\mu$.
As seen from the behavior of the effective potential
in Fig.~\ref{fig:3},
a second-order transition occurs in the range $T=$ 100 -- 120 MeV
in the cases with $\mu=0$ and $\mu=100$ MeV.
At $\mu=300$ MeV, however, a first-order transition occurs 
between $T=75$ MeV and 80 MeV. 
Similarly, in Fig.~\ref{fig:4}, a first-order transition is seen to occur 
when the chemical potential is changed between 
$\mu=280$ MeV and 290 MeV at low temperature ($T=80$ MeV),
while the transition is second order at fixed $T=100$ MeV.
In Fig.~\ref{fig:5}, the quark number density $\rho$ is plotted 
in units of the normal nuclear matter 
density in terms of quark numbers, 
$\rho_0=3\times 0.17\ {\rm fm}^{-3}$, as a function 
of $\mu$ with fixed $T=80$ MeV and 100 MeV. 
The transition points are $\mu=282$ MeV at $T=80$ MeV and 
$\mu=187$ MeV at $T=100$ MeV.
In the former case, the quark number density has a gap.
In the latter, 
it is continuous but its derivative (the susceptibility) has a gap.

In Fig.~\ref{fig:6}, we plot the contour map of the Landau potential 
${\ov V}(\langle{\bar \psi}\psi\rangle,\rho; \mu, T)$ given in (\ref{32}) 
as a function of the quark condensate and the quark number density
at several values of $\mu$ with fixed $T$. 
Note that a tri-critical point is located at
$(T_t, \mu_t)\simeq (97 , 203)$ MeV. 
The vertical axis represents 
$\langle{\bar \psi}\psi\rangle/\Lambda_{\rm QCD}^3$, and the horizontal 
axis represents the quark number density in units of $\rho_0$. 
We plot 
the contour map at $T=100$ MeV,
where a second-order transition occurs, in the left panels of Fig.\ref{fig:6}. 
At $\mu=0$, the two minima are positioned symmetrically 
in the $\rho$-$\langle{\bar \psi}\psi\rangle$ plane at $\rho=0$ with
finite values of the quark condensate, $\langle{\bar \psi}\psi\rangle$. 
As the chemical potential increases, the two minima approach each other 
for finite $\rho$, and then fuse continuously to form a single minimum
at $\langle{\bar \psi}\psi\rangle=0$ and finite $\rho$.

The contour map for $T=80$ MeV is plotted in the right panels 
of Fig.\ref{fig:6}. 
At $\mu=0$, the two minima are positioned symmetrically 
in the $\rho$--$\langle{\bar \psi}\psi\rangle$ plane at $\rho=0$ and 
finite values of the quark condensate, $\langle{\bar \psi}\psi\rangle$, 
just as in the $T=100$ MeV case. 
By contrast, however,  one new local minimum appears 
at $\langle{\bar \psi}\psi\rangle=0$ as $\mu$ increases, 
and three local minima exist in a certain range of $\mu$. 
At the critical point, these three minima are energetically degenerate,
and we observe a first-order chiral phase transition.

\section{Summary}

We have analyzed the chiral phase transition of a QCD-like theory 
at finite temperature and density using the effective potential.
We have devised a method to derive an effective potential in order to 
elucidate the global behavior of the phase transition. 
There we introduced a bilocal external source field and solved the 
Schwinger-Dyson equation with this source field. 
This solution of the SD equation gives the extremum of the 
effective potential with the source field. 
Then, subtracting the interaction energy with the source 
we obtained the original effective potential for this configuration.
In this way we showed the shape of the effective potential of the QCD-like 
theory at finite temperature, $T$, and 
quark chemical potential, $\mu$. 
From the temperature dependence of the effective potential, 
we have a second-order phase transition along the $\mu=0$ line. 
Contrastingly, we have a first-order phase transition 
along the $T=0$ line. 
Also, we have introduced the Landau potential and plotted its contour 
map with respect to the quark condensate and the quark number density. 
We conclude that our method to construct the effective potential of
the QCD-like theory is very useful for understanding 
the features of the chiral phase transition of the model.

\section*{Acknowledgements}
The authors would like to thank M.~Harada for bringing 
Ref.~\citen{RWH91} to their attention.
This work was partially
supported by Grants-in-Aid from the Japanese Ministry of Education, 
Culture, Sports, Science and Technology, [Nos.15740156 (Y.T.), 
13440067 (H.F) and 16740132 (H.F.)].

\appendix
\section{Calculation of the Effective Potential in the NJL Model}

For pedagogical reasons, 
here we apply our method with a source field coupled 
quadratically to the 
auxiliary field, to the Nambu-Jona-Lasinio (NJL) model, 
whose Lagrangian density takes the form 
\begin{equation}\label{a1}
\mathcal{L}=\ov{\psi}  i\slpartial \psi
+\frac{\lambda}{2N}\left((\ov{\psi}\psi)^2
+(\ov{\psi}i\gamma_5{\boldmath \tau}\psi)^2\right) .
\end{equation}
We introduce the auxiliary field into the scalar part of the 4-fermi
interaction and ignore the pseudo-scalar part for the moment, 
as we know that the potential is chirally symmetric.
To leading order in the $1/N$ expansion, 
we obtain the effective potential at finite $T$ and $\mu$ as
\begin{eqnarray}\label{a3}
V(T,\mu,\chi)
&=&\!\!
\frac{\chi^2}{2\lambda}
-2N_cN_fT\!\sum^\infty_{n=-\infty}\!\int\frac{d^3\bk}{(2\pi)^3} 
\textrm{ln}[(\omega_n-i\mu)^2+\bk^2+\chi^2] \nonumber\\ 
&=&
\frac{\chi^2}{2\lambda}-\frac{N_cN_f}{2\pi^2} 
\int^{\chi^2}_0 dx \int^{\Lambda}_0 d|\bk| \frac{\bk^2}{\sqrt{\bk^2+x}} 
(1-n_+(x)-n_-(x)) ,
\end{eqnarray}
where $\chi=-\lambda\ov{\psi}\psi/N$ is an auxiliary field,
and  $n_\pm(x)=[\exp(\sqrt{\bk^2+x}\mp\mu)+1]^{-1}$.

\begin{figure}[t]
\begin{center}
\includegraphics[height=4.8cm]{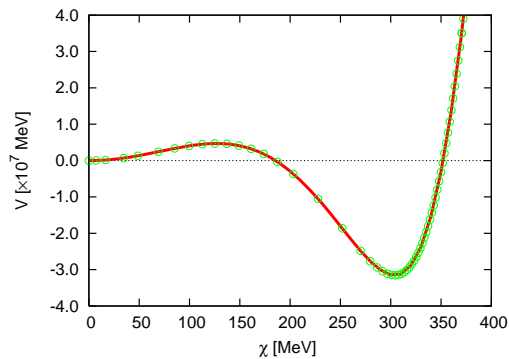} 
\end{center}
\caption{Two effective potentials $V(\chi)-V(\chi=0)$ (solid curve) 
and ${V}(\chi,J)-V(\chi=0,J=0)$ (round points) composed 
in the case $T=10$ MeV and $\mu=320$ MeV. 
}
\label{fig:8}
\end{figure}

The functional form of the potential is easily obtained through the 
numerical integration of Eq.~(\ref{a3}). 
To demonstrate our method,
we consider the case of a constant source coupled to the field
$\chi^2$:\cite{IST95} 
\begin{equation}\label{a5}
\widetilde{V}(\chi,J)=V(\chi)+\frac{1}{2}J\chi^2 .
\end{equation}
Rewriting the source $J$ as $J=c\lambda^{-1}$,
we obtain 
\begin{eqnarray}\label{a6}
\widetilde{V}(\chi,J) &=&
(1+c)\frac{\chi^2}{2\lambda}-\frac{N_cN_f}{2\pi^2} 
\int^{\chi^2}_0\!\! dx \int^{\Lambda}_0\! d|\bk| \frac{\bk^2}{\sqrt{\bk^2+x}} 
(1-n_+(x)-n_-(x)) ,\qquad
\end{eqnarray}
and the corresponding SD equation, $\partial \widetilde{V}/\partial \chi=0$: 
\begin{eqnarray}\label{a7}
\frac{\chi_c}{\lambda}&=&
\frac{1}{1+c}\frac{N_cN_f\chi_c}{\pi^2} \int^{\Lambda}_0 d|\bk| 
\frac{\bk^2}{\sqrt{\bk^2+\chi_c^2}} 
(1-n_+(\chi_c^2)-n_-(\chi_c^2)) . 
\end{eqnarray}
The value of the effective potential at $\chi=\chi_c$ is obtained
from $\tilde V_{\rm ex}(\chi_c)$ using the Legendre transformation:
\begin{eqnarray}\label{a8}
\bar V(\chi_c)=\tilde V_{\rm ex}(\chi_c)-\frac{1}{2}J \chi_c^2 \ .
\end{eqnarray}
In the stationary-phase approximation for $\chi$, we should have the trivial 
result
\begin{eqnarray}
V(\chi)=\bar V(\chi).
\end{eqnarray}
One of the advantages of the source (\ref{a5}) is that 
it allows us to probe the
non-convex effective potential near a second-order transition.
Near a first-order transition point, the SD equation (\ref{a7}) has
two non-trivial solutions for a certain range of $c$. 
We need both the solutions
in order to re-construct the full shape of the potential via Eq.~(\ref{a8}).

We present the result in the massless two-flavor case
($N_c=3$ and $N_f=2$) here.
We fixed the three-momentum cutoff as $\Lambda \simeq 653$ MeV 
and the coupling constant as $\lambda\Lambda^2 \simeq 4.29$
with the constituent quark mass $313$ MeV and $f_\pi=93$ MeV. 
In Fig.\ref{fig:8}, 
we compare the effective potential $\bar V(\chi)$ (circles),
constructed using the applied source and the 
Legendre transformation, with the original $V(\chi)$ (solid curve).
It is seen that they are essentially coincident, as should be the case. 
Hence, we conclude that our method
successfully reproduces the potential in the non-convex region\cite{IST95}.


\begin{thebibliography}{99}
\bibitem{CP75}
J.~C.~Collins and M.~J.~Perry, Phys.~Rev.~Lett. {\bf 34} (1975), 1353.
\bibitem{QGP}
See, for example, {\it Quark-Gluon Plasma 3}, ed. R. C. Hwa and X.-N. Wang 
(World Scientific, Singapore, 2004). 
\bibitem{KMTT93}
See, for example, T. Kunihiro, T. Muto, T. Takatsuka, R. Tamagaki 
and T. Tatsumi, Prog. Theor. Phys. Suppl. No. 112 (1993). 
\bibitem{LQCDatNara}
See, for example, 
Prog. Theor. Phys. Suppl. No. 153 (2004). 
\bibitem{HK94}
T. Hatsuda and T. Kunihiro, Phys. Rep. {\bf 247} (1994), 1.\\
S. P. Klevansky, Rev. Mod. Phys. {\bf 63} (1992), 649. 
\bibitem{AY89}
M. Asakawa and K. Yazaki, Nucl. Phys. A {\bf 504} (1989), 668. 
\bibitem{BCCGP89}
A. Barducci, R. Casalbuoni, S. De Curtis, R. Gatto and G.~Pettini, 
Phys. Lett. B {\bf 231} (1989), 463.
\bibitem{BCP94}
A Barducci, R. Casalbuoni, G. Pettini and R.~Gatto, Phys. Rev. D {\bf 49}
(1994), 426; ibid. {\bf 41} (1990), 1610. 
%
\bibitem{BL}
D. Bailin and A. Love, Phys. Rep. {\bf 107} (1984), 325. 
\bibitem{II}
M. Iwasaki and T. Iwado, Phys. Lett. {\bf 350} (1995), 163; 
Prog. Theor. Phys. {\bf 94} (1995), 1073. 
\bibitem{RW}
M. Alford, K. Rajagopal and F. Wilczek, Phys. Lett. B {\bf 422} (1998), 247.
\bibitem{RSSV}
R. Rapp, T. Sch\"afer, E. V. Shuryak and M. Velkovsky, Phys. Rev. Lett. 
{\bf 81} (1998), 53. 
\bibitem{ABKMN90}
K.-I. Aoki, M. Bando, T. Kugo, M. G. Mitchard and H. Nakatani, 
Prog. Theor. Phys. {\bf 84} (1990), 683. 
\bibitem{K92}
T. Kugo, 
{\it Dynamical Symmetry Breaking}, ed. by K. Yamawaki 
(World Scientific, Singapore, 1992), p.35. 
\bibitem{H91}
K. Higashijima, Prog. Theor. Phys. Suppl. No. 104 (1991), 1. 
\bibitem{FT95}
H. Fujii and Y. Tsue, Phys. Lett. B {\bf 357} (1995), 199. 
\bibitem{TY97}
Y. Taniguchi and Y. Yoshida, Phys. Rev. D {\bf 55} (1997), 2283. 
\bibitem{HS98}
M. Harada and A. Shibata, Phys. Rev. D {\bf 59} (1998), 014010. 
\bibitem{KMT00}
O. Kiriyama, M. Maruyama and F. Takagi, Phys. Rev. D {\bf 62} (2000), 105008. 
\bibitem{I02}
T. Ikeda, Prog. Theor. Phys. {\bf 107} (2002), 403. 
\bibitem{T03}
S. Takagi, Prog. Theor. Phys. {\bf 109} (2003), 233.
\bibitem{A03}
H. Abuki, Prog. Theor. Phys. {\bf 110} (2003), 937. 
\bibitem{CJT74}
J. M. Cornwall, R. Jackiw and E. Tomboulis, Phys. Rev. D {\bf 10} (1974), 2428.
\bibitem{RWH91}
R. W. Haymaker, Riv. Nuovo. Cim. {\bf 14} (1991), 1.
\bibitem{HM84}
K. Higashijima, Phys. Rev. D {\bf 29} (1984), 1228.\\
V. A. Miransky, Sov. J. Nucl. Phys. {\bf 38} (1983), 280. 
\bibitem{sasaki}
S.~Sasaki, H.~Suganuma and H.~Toki, Phys.~Lett.~B {\bf 387} (1996), 145.
\bibitem{MN74}
T. Maskawa and H. Nakajima, Prog. Theor. Phys. {\bf 52} (1974), 1326. 
\bibitem{pagels}
H.~Pagels and S.~Stokar, Phys.~Rev.~D {\bf 20} (1979), 11.
%
\bibitem{KKZP04}
O. Kaczmarek, F. Karsch, F. Zantow and P. Petreczky,
Phys.~Rev.~D {\bf 70} (2004), 074505.
\bibitem{morozumi}
T.~Morozumi and H. So, Prog. Theor. Phys. {\bf 77} (1987), 1434. 
\bibitem{FO04}
H.~Fujii and M.~Ohtani, Phys.~Rev.~D {\bf 70} (2004), 014016.
\bibitem{NG92}
N.~Goldenfeld, {\it Lectures on Phase Transitions and the
        Renormalization Group} (Addison-Wesley Pub.\ Co., 1992). 
\bibitem{IST95}
H.~Ichie, H.~Suganuma and H.~Toki, Phys.~Rev. D {\bf 52} (1995), 2944.
\end{thebibliography}
\end{document}